\documentclass{aa}%[referee]
\usepackage{natbib}
\usepackage{graphicx}
\usepackage[mathcal]{eucal}
\usepackage{amssymb} 

\newcommand{\ds}{\displaystyle}
\newcommand{\inv} {\frac {1}}

\long\def\jumpover#1{{}}

\newcommand{\deriv} [2] {\frac {d #1 } {d #2} }

\newcommand{\eqn} [1] {
\begin{equation}#1
\end{equation}}
\newcommand{\eqna} [1] {
\begin{eqnarray}#1
\end{eqnarray}}

\newlength{\lenA} % 
\begin{document}

%\title{Non-Gaussian character of stochastic excitation of solar p~modes.}
\title{Numerical 3D constraints on convective eddy time-correlations : consequences for  stochastic excitation of solar p~modes.}
\author{Samadi R. \inst{1,2} \and  Nordlund {\AA}. \inst{3} \and Stein R.F. \inst{4} \and Goupil M.J.\inst{2}   \and  Roxburgh I.  \inst{1,2} }

\institute{
Astronomy Unit, Queen Mary, University of London, London E14NS, UK \and
Observatoire de Paris, LESIA, CNRS UMR 8109, 92195 Meudon, France \and 
Niels Bohr Institute for Astronomy, Physics, and Geophysics, Copenhagen, Denmark\and
Department of Physics and Astronomy, Michigan State University, Lansing, USA%\and Institute of Astronomy, University of Cambridge, Cambridge CB3 0HA, UK 
}
\offprints{R. Samadi}
\mail{Reza.Samadi@obspm.fr}
\date{\today} % Received / Accepted}

\titlerunning{Numerical 3D constraints on convective eddies time-correlations.}

%%%%%%%%%%%%%%%%%%%
\abstract{
A 3D simulation of the upper part of the solar convective zone is used to
 obtain information on the frequency component, $\chi_k$, of the 
correlation product of the turbulent velocity field. This component plays an important role in  the stochastic
excitation of acoustic oscillations. % of \citet{Samadi00I}.
A time analysis of the solar simulation shows that a gaussian 
function does not correctly reproduce the $\nu$-dependency of $\chi_k$ 
inferred from   the 3D simuation  in the frequency range where the
acoustic energy  injected into the solar p~modes is important  ($\nu \simeq 2 - 4$~mHz).
The  $\nu$-dependency of  $\chi_k$ is fitted with different  analytical  functions  which can then conveniently be used to compute the
acoustic energy supply rate $P$ injected into the solar radial oscillations.
With constraints from a 3D simulation, adjustment of free parameters to solar
data is no
longer necessary and is not performed here.  The result is compared with solar seismic data.
Computed values  of $P$ obtained with the analytical function which  fits best  
$\chi_k$ are found   $\sim 2.7$  times larger than those obtained with the gaussian model
and reproduce better the solar seismic observations. 
This non-gaussian description  also leads to a Reynolds stress contribution of the
same order  as the one arising from  the advection of the turbulent
fluctuations  of entropy by the turbulent motions.  Some discrepancy between
observed and computed $P$ values still exist at high frequency and possible
causes for this discrepancy  are discussed.
\keywords{convection - turbulence - stars:oscillations - Sun:oscillations}
 }
\maketitle

%%%%%%%%%%%%%%%%%%%%%%%%%%%%%%%%%%%%%%%%%%
\section{Introduction}

Solar oscillations are believed to be stochastically excited by turbulent 
convection in the outer part of the Sun. The excitation is caused 
by turbulent convective motions  which generate acoustic energy 
which in turn is injected into the p~modes.

Models of stochastic excitation of stellar p~modes have been  
proposed by several authors, \citep[e.g.][]{GK77,Osaki90,Balmforth92c,GMK94}.
These models use simplified models to describe the  dynamics  of the turbulent medium.
For instance these approaches \citep{GK77,Balmforth92c} assume a gaussian 
function for representing $\chi_k$, the frequency component of the  correlation product  of the turbulent velocity field.
As pointed out by \citet{Samadi01}, the way the  component 
  $\chi_k$   is modeled plays a crucial role in 
controlling the extent of the excitation region of a given mode 
and hence the total amount of acoustic energy injected into the mode. 
In the following, $\chi_k$  will also be referred to as \emph{the dymamic model of turbulence} and 
 \emph{dynamic} will refer to time-dependence or frequency-dependence.

Direct computations of the rate at which the solar p~modes are excited have been performed by 
\citet{Stein01II} using 3D simulations of the upper part of the solar convective zone.
They found good agreement between their numerical results  and the solar seismic observations.
This direct but time consuming approach did not address the role of the 
dynamic  properties of the turbulent medium on the excitation mechanism.

In contrast semi-analytical formulations for $P(\nu)$ offer the advantage 
of testing {\it separately} several properties entering the excitation mechanism.
Here we consider the formulation by \citet[ hereafter Paper~I, see also
\citet{Samadi01} for  a  summary]{Samadi00I}   
which includes a detailed  treatment of the \emph{time averaged} 
and \emph{dynamic} properties of the turbulent convective medium.

The impact of the averaged properties have been investigated by \citet{Samadi02I}.
The authors used a 3D simulation of the upper part of the solar convective zone
to constrain the averaged  properties of the turbulent convective medium.
The computed  rates $P$ at which the solar p~modes are excited were found
to be larger than those computed with a 1D  mixing-length  solar model
 but still \emph{underestimate} the solar seismic data by a factor $\sim 2.5$. 
It was also found that the Reynolds tensor contributes 
about $20\,\%$  of the total acoustic energy injected into the solar p~modes,
in contrast with direct 3D estimations \citep{Stein01II}.
These discrepancies were attributed to  the assumed gaussian function  for the 
dynamic model of turbulence.
%the frequency dependence of the correlation product of the turbulent velocity field.

In the present paper we therefore derive an empirical dynamic model of turbulence
obtained from a 3D simulation
of the upper part of the solar convective zone  and then 
study the consequences  
of using this model on the computed excitation rates $P$.
We compare our computation with solar seismic data and 
 finally obtain on  an improved model of stochastic excitation.

The paper is organised as follows:  The basic theoretical background and notations    are recalled in Sect.~2. 
In Sect.~3, a 3D simulation of the upper part of the solar convective zone is used to characterise  $\chi_k$   in the domain where stochastic excitation takes place. 
The inferred $\nu$-dependency of  $\chi_k$  is compared with the gaussian function and fitted with different   non-gaussian functions.
These functions are then used in Sect.~4  to compute the excitation rate $P$ for radial p~modes.
The results are compared with solar seismic observations as provided by \citet{Chaplin98} 
and with computations in which the gaussian function is assumed. Sect.~5 is dedicated to discussions and conclusions.

%%%%%%%%%%%%%%%%%%%%%%%%%%%%%%%%%%%%%%%%%%
\section{Stochastic excitation}
\label{sec:Stochastic excitation}

%-----------------------------------------
\subsection{The  model of stochastic excitation}

We consider the model of stochastic excitation as described  in Paper~I and
 assume here -~as  in \citet{Samadi02I}~- that  injection of acoustic energy into 
the modes is  isotropic and consider only radial p~modes.
Accordingly, the rate at which a given mode with frequency $\omega_0$ is excited can be written  as~:
\eqna{
P(\omega_0)  & = &  \frac{ \pi^{3}  }
{ 2   I}  \,    
\int_{0}^{M}{\rm d}m \,  {\Phi \over 3}  \, \rho_0 w ^4 \, 
\left \{ {16\over 15}  \, \frac{\Phi}{3} \, \left (\deriv { \xi_{\rm r}} {r} \right )^2 \,  S_R   \right .   \nonumber 
\\ & &  \hspace{1cm} \left .  \;   + \,   {4\over 3}  \,
\left ( \frac{\alpha_s \, \tilde s}{\rho_0 \, w} \right )^2  \,   \frac{g_{\rm r}  }{\omega_0^2} \,   S_S\right \} \; .
\label{eqn:P}
} % 
In Eq.~(\ref{eqn:P}), $\rho_0$ is the mean density, $\displaystyle{\xi_{\rm r}}$ is the
 radial component of the fluid displacement adiabatic eigenfunction ${\bf
 \xi}$,   $I$ is the mode inertia (Eq.~(\ref{eqn:I})),  
$\displaystyle{\alpha_s =\left ( \partial p /\partial s  \right )_\rho}$ 
where $p$ denotes the  pressure and $s$ the entropy,
 $\tilde s$ is the rms value of the entropy fluctuations  which are assumed
 to arise solely from turbulence,  
$g_r(\xi_{\rm r},r)$ is a function involving the first 
and the second derivatives of  ${\xi}_r$ with respect to $r$, 
$\Phi$ is a mean anisotropy factor defined by \citet{Gough77} as  
\eqn{
\Phi(r) \equiv 
\frac{\overline{  <{\bf u}^2> - <{\bf u}>^2}}{w^2(r)} 
\label{eqn:phiz}
}
where $\bf u$ is the velocity field, $< . >$ denotes  horizontal  average,  $\overline{()}$ denotes time average,  and $w(r)$  is the mean vertical 
velocity ($\ds w^2 \equiv \overline{<u_z^2>-<u_z>^2}$). Expressions for  $g_r(\xi_{\rm r},r)$
 are  given in \citet{Samadi02I}.

The driving sources  $S_R(r,\omega_0)$ and $S_S(r,\omega_0)$ 
arise from the Reynolds stress and the entropy fluctuations respectively ~:
\eqna{
S_{R}(r,\omega_0) & = & \int_0^\infty { {\rm d} k \over  k^2} \,
  \frac{E(k,r)}{u_0^2} \, \frac{E(k,r)}{u_0^2} \, \chi_k ( \omega_0 , r) \label{eqn:SR} \\
S_{S}(r,\omega_0) & = &  \int_0^\infty { {\rm d} k \over  k^2} \,
  \frac{E(k,r)}{u_0^2} \,  \frac{E_s(k,r)}{\tilde s^2} 
\, \nonumber \\ & & \times \,\int_{-\infty}^{+\infty}{\rm d} 
\omega \, \chi_k ( \omega_0 + \omega,r)\chi_k (\omega,r) \label{eqn:SS}
}
where    $u_0(r) \equiv \sqrt{ \Phi/3}   \,  w $  is introduced for convenience, 
$E(k,r)$ is the time averaged  turbulent kinetic energy spectrum, $E_s(k,r)$ is the time 
averaged turbulent spectrum associated with the entropy fluctuations and  $\chi_k(\omega,r)$ 
is the frequency-dependent part of the correlation product of the turbulent velocity field 
(see Section~\ref{sec:The dynamic model of turbulence}). In order to simplify the notation, 
we drop the explicit $r$ dependence of the quantities in 
Eqs.~(\ref{eqn:P}-\ref{eqn:SS}).

%-----------------------------------------
\subsection{The dynamic model of turbulence}
\label{sec:The dynamic model of turbulence}

The dynamic model of turbulence is represented by $\chi_k(\omega)$.  
In order to give a  precise meaning to $\chi_k(\omega)$,  we recall first some 
theoretical relations.
Excitation by  Reynolds stresses involves  $\phi_{i,j}( k,\omega)$,  
the Fourier transform of the second-order velocity correlations; 
here the indices $i$ and $j$ refer to any of the 3 directions of the
velocity field. For incompressible, homogeneous and isotropic turbulence, 
$\phi_{ij}( k,\omega)$  has the form \citep{Batchelor70} :      
\eqna{
\phi_{ij}(  k,\omega) &  =& \frac { E( k,\omega) } { 4 \pi k^2}  \left( \delta_{ij}- \frac {k_i k_j} {k^2}  \right)
\label{eqn:phi_ij}
} 
where $ E( k,\omega) $ is the turbulent kinetic energy spectrum as a function of $k$ and $\omega$ and $\delta_{ij}$ is the Kronecker tensor.
 Following  \citet{Stein67}, $ E(k,\omega) $ is decomposed  as
\eqna{
E( k,\omega) =E(  k) \, \chi_k(\omega)
\label{eqn:ek_chik}
}
where $\chi_k(\omega)$ satisfies the normalisation condition \citep[Chap 8.1]{Tennekes82}~:
\eqn{
\int_{-\infty}^{+\infty} d\omega \,  \chi_k (\omega)  = 1 \; .
\label{eqn:chi_omega_norm}
}
According to the decomposition of Eq.~(\ref{eqn:ek_chik}) , 
$\chi_k(\omega) $ is -~at fixed $k$~- the frequency component of $E(k,\omega)$.

\vspace{0.2cm}

According to Eq.~(\ref{eqn:phi_ij}) and (\ref{eqn:ek_chik}), $\chi_k(\omega) $   then
represents   the frequency dependence of  $\phi_{i,j}(\vec k,\omega)$.
In other words, $\chi_k(\omega) $  measures -~  in the frequency and $k$
wavenumber  domains ~-  the evolution of the velocity correlation
 between two distant points of the turbulent medium.

The same decomposition of Eq.~(\ref{eqn:chi_omega_norm}) is assumed for $E_s(k,
\omega)$. 
This leads to introducing $\ds \chi_k^s$, the frequency-dependent 
part of the correlation product of the entropy fluctuation.  
For simplifying the  computation of $P$,  as in Paper~I, we assume   $\ds
\chi_k^s = \chi_k$. 
We have checked  that $\chi_k^s$ and  
$ \chi_k$ have almost the same behaviour in the region where 
excitation by entropy fluctuations is significant.

\vspace{0.2cm}

In the present work, we consider only the excitation of radial p~modes.
Let  $E_z(k,\omega)$ be  the vertical component of the kinetic energy spectrum.
We consider that $E_z(k,\omega)$ can be decomposed 
as   $E(k,\omega)$ (Eq.~(\ref{eqn:chi_omega_norm})).
For isotropic turbulence we then have  $\ds E(k,\omega)=3 \,
E_z(k,\omega)$,  $E(k)= 3 \, E_z(k)$  and $\ds \chi_k^z = \chi_k$, which is
equivalent to stating that the averaged and dynamic properties  of the  velocity field are
the  same in all 3 directions.

%The study by \citet{Samadi02I} concerned the time and spatial averaged properties of the turbulent medium in which the excitation occurs.
The anisotropy factor  $\Phi$ -~ introduced in the expression for $P$,
Eq.~(\ref{eqn:P}) ~- 
partially takes into account the 
spatial and temporal 
anisotropy of the turbulence  ($\Phi=3$ corresponds to a isotropic turbulence).
It has been found in \citet{Samadi02I} that $\Phi \simeq 2$ within the region where most of the excitation occurs.
This shows that the time and space averaged properties of the 
medium are indeed anisotropic.
One can therefore expect that the dynamic properties of turbulence differ 
between the horizontal and vertical directions.
 As the excitation of radial p~modes is predominantly governed by turbulent
elements  moving in the vertical direction, an open question is whether one
should consider $\ds \chi_k^z$ rather than $ \chi_k$ in Eq.~(\ref{eqn:SR}) and
Eq.~(\ref{eqn:SS}) when taking into account the dynamic 
properties of the turbulence.
In the present work, we therefore characterise both $\ds \chi_k^z$ and  
$ \chi_k$ from a 3D simulation and assess the consequences of using 
either $\ds \chi_k^z$ or $ \chi_k$ in the calculation of $P$.

%---------------------------------
\subsection{A Gaussian function  for $\chi_k$}

\citet{Stein67} and  \citet{Musielak94} suggested several analytical forms for $\chi_k(\omega) $.
The  Gaussian Function (GF hereafter) is the simplest choice and is defined as
\eqn{
\chi_k (\omega ) = \inv  { \omega_k \, \sqrt{\pi}}  e^{-(\omega / \omega_k)^2} 
\label{eqn:GF}
}
where $\omega_k$ is its linewidth.

In the time domain, the gaussian function, Eq.~(\ref{eqn:GF}),  is 
the Fourier transform of  a  gaussian function 
whose  linewidth is  equal to  $2 \tau_k$, where 
 $\tau_k$ is a characteristic time correlation 
length. 
Hence $\omega_k$ and $\tau_k$ are related to each other as
\eqn{
\omega_k= {2  \over  \tau_k} \; .
\label{eqn:omegak}
} 
The characteristic time $\tau_k$ is usually associated with the characteristic 
correlation time-scale of an eddy with wavenumber $k$. 
As in \citet{Balmforth92c},  we define it as 
\eqn{
\tau_k \equiv {\lambda  \over k u_k}
\label{eqn:tauk}
} 
where the velocity  $u_k$ of the eddy with wave number $k$ 
is related to the kinetic  energy spectrum $E(k)$ by   \citep{Stein67} 
\eqn{
u_k^2 =  \int_k^{2 k}  dk \, E(k) \; .
\label{eqn:uk2}
}

The parameter $\lambda$ in Eq.~(\ref{eqn:tauk}) 
accounts for our lack of precise  knowledge of the
 time correlation  $\tau_k$ under stellar conditions.

In the calculation of $P$, a GF is  usually assumed for $\chi_k$ 
(e.g. \citet{GK77}, \citet{Balmforth92c}).
This assumption is equivalent to supposing that two distant points in the 
turbulent medium are uncorrelated.

In Sect.~\ref{sec:Constraints from the 3D simulation}, we use a 3D simulation
of  the upper part of the solar convective zone to derive the $\nu$-dependencies
 of $\ds \chi_k^z$ and $\chi_k$.
Inferred   $\nu$-dependencies of  $\ds \chi_k^z$ and $\chi_k$ are compared to that of the GF.
We next determine several analytical forms for  $\ds \chi_k^z$ and $\chi_k$ that can better 
represent their $\nu$-dependencies.

%We assume  for the sake of simplicity that these analytical forms remain constant with depth.
%However as $\tau_k$ and thus $\omega_k$ varies rapidly with depth,  these  analytical forms  will be expressed as a function of the unidimensional quantity $\omega / \omega_k$ where $\omega_k$ varies with depth according to Eqs.(\ref{eqn:omegak}-\ref{eqn:uk2}).

%%%%%%%%%%%%%%%%%%%%%%%%%%%%%%%%%%%%%%%%%%
\section{Constraints from the 3D simulation}
\label{sec:Constraints from the 3D simulation}

The analysis of a 3D simulation of the upper part of the solar convective
zone  provides  constraints for several physical parameters 
that enter   the theoretical expression for the energy supply 
rate $P$ injected into the solar p~modes (Eq.~\ref{eqn:P}).
The constraints may be considered to be of two types : 
\begin{itemize}
\item[$\bullet$] \emph{static} constraints (\emph{static} refers to 
spatial and time averages)  determine the actual wavenumber dependency of
$E(k,z)$, the kinetic turbulent spectrum,  
and $E_s(k,z)$, the turbulent  spectrum associated with the entropy.
The \emph{static} constraints also  determine the  depth profile  of the wavenumber $k_0^E$  at which   convective energy is injected into the
turbulent inertial range of $E$ (as in \citet{Samadi02I} we assume that the wavenumber  $k_0^{E_s}$, at which   convective energy is injected into the
turbulent inertial range of $E_s$, is equal to $k_0^E$).
They also provide the depth dependence of $u^2$, the  mean square velocity,
$w^2$, the mean square vertical component of the velocity, $\tilde s^2$, 
the mean square values of  entropy fluctuations  and $\Phi=u^2
/w^2$, the mean values of the anisotropy studied in \citet{Samadi02I}. 
\item[$\bullet$]The \emph{dynamic}  constraints,  on the other hand, concern 
the  frequency component $\chi_k$ and $\ds \chi_k^z$ (see Sect.~\ref{sec:The dynamic model of turbulence}). 
\end{itemize}

The static constraints have been   established in \citet{Samadi02I}.
Here we investigate the dynamic constraints.

\subsection{The 3D simulation}

We study a 3D simulation of the upper part of the solar convection zone
obtained with the 3D numerical code developed by Stein \& Nordlund (1998).

The simulated domain is 3.2 Mm deep and its surface is 6 x 6 ${\rm Mm}^2$. 
The grid of mesh points is 256 x 256 x 163 (i.e. $\sim 23 {\rm km} \times 23 
{\rm km} \times  37 {\rm km}$), the total duration 27 mn and the sampling time 30s.

Outputs of the simulation considered   in \citet{Samadi02I} are 
 the velocity field  $\vec u(x,y,z,t)$ and the entropy $s(x,y,z,t)$ 
 where -~as in \citet{Samadi02I}~-  $z=r-R_\odot$ and  $R_\odot$ is 
the radius at the photosphere (i.e. where $T=T_{\rm eff}$). 
The quantities $\vec u(x,y,z,t)$ and  $s(x,y,z,t)$  were used to 
determine the quantities $E(k,z)$, $E_s(k,z)$, $w$ and $\tilde s^2$ 
involved in the theoretical expression for the excitation rate $P$.
In the present work we use  the velocity field  $\vec u(x,y,z,t)$ 
to characterise $\chi_k$ and $\ds \chi_k^z$.

%-----------------------------------------------
\subsection{Fourier transform}

 At four different layers of the simulated domain, we 
compute the 3D Fourier transform, with respect to time and in the 
horizontal plane,  of the velocity field ${\bf u}$. 
These layers cover a region where modes with frequency $\nu \gtrsim 2$~mHz 
are predominantly excited.

This provides $\vec {\hat u}(\vec k,z,\nu)$  where  $\vec k$ is the wavenumber in the horizontal plane.
Next we integrate $\vec {\hat u}^2(\vec k,z,\nu)$  over circles 
with radius $k=\| \vec k \|$ at each given layer $z$. 
This yields $ {\hat {\vec u}}(k,z,\nu)$ and therefore $E (k,\nu,z) \equiv  {\hat {\vec u}}^2(k,\nu,z)$.
The quantity $\chi_k(\nu,z)$ is  the frequency component of $E (k,\nu,z)$, at fixed $k$.
Hence, according to  Eq.(\ref{eqn:ek_chik}) and (\ref{eqn:chi_omega_norm}),
$\chi_k(\nu,z)$ is obtained  from $E (k,\nu,z)$ as: 
\eqn{
\chi_k(\nu,z) =\frac{ E (k,\nu,z)} { \ds \int {\rm d} \nu \, E (k,\nu,z)}
} %%_{-\nu_{\rm max}}^{\nu_{\rm max}}
 where the integration over $\nu$ is performed over the frequency range 
 $[-\nu_{\rm max},\nu_{\rm max}]$ corresponding to the window of the Fourier analysis 
 with respect to time ($\nu_{\rm max}\simeq 16$~mHz).  

We proceed in the same manner for the vertical component of $\vec u$. 
This then provides  $E_z(k,\nu,z) \equiv  {\hat { u}_z}^2(k,\nu,z)$ and $\ds \chi_k^z(\nu,z)$.

%-----------------------------------------------
\subsection{Inferred properties of $\chi_k$ and $\ds \chi_k^z$}
\label{sec:Inferred properties of chikz}

\begin{figure*}[ht]
\begin{center}
        \resizebox{\lenA}{!}{\includegraphics  {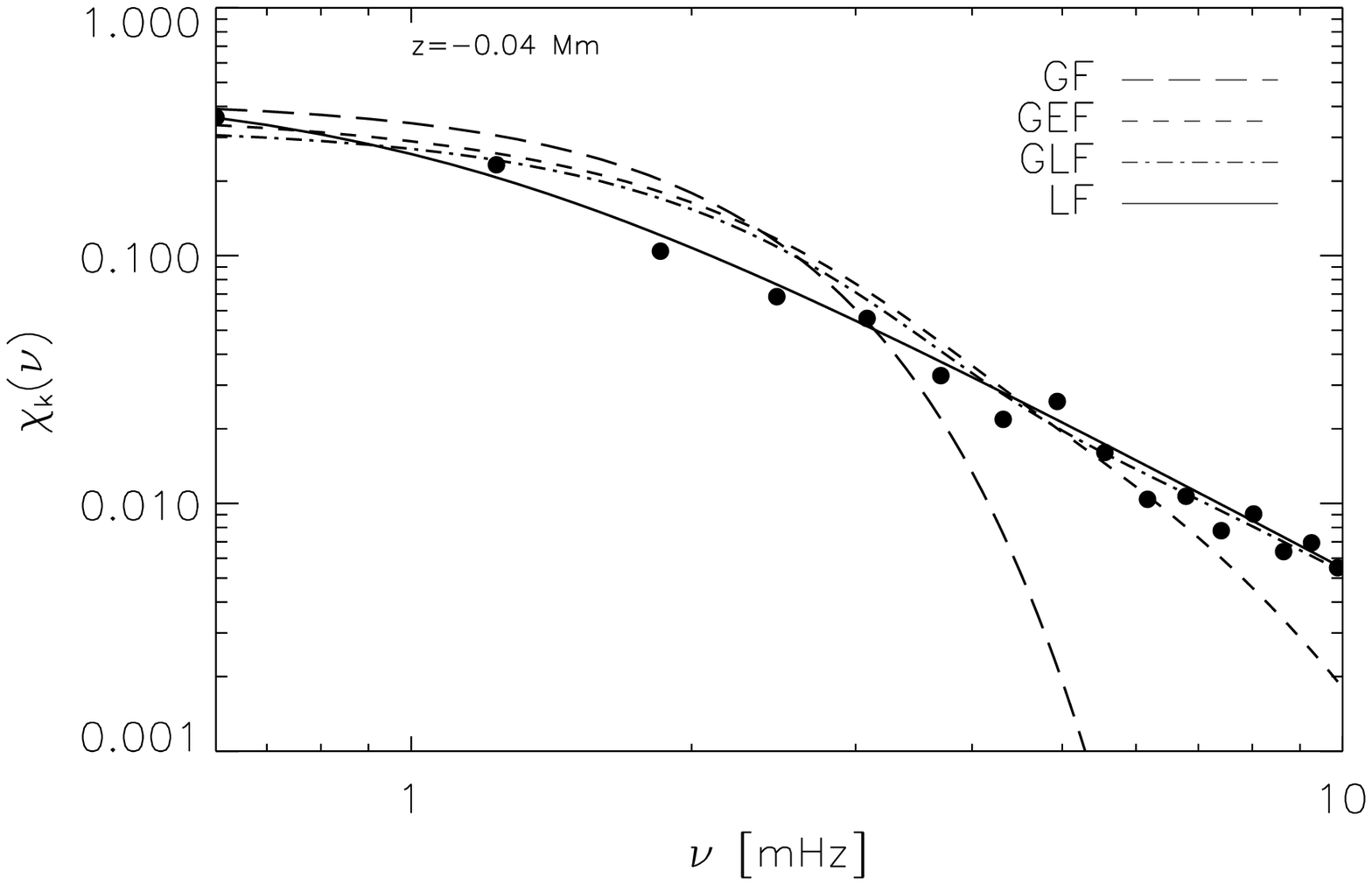}}
        \resizebox{\lenA}{!}{\includegraphics  {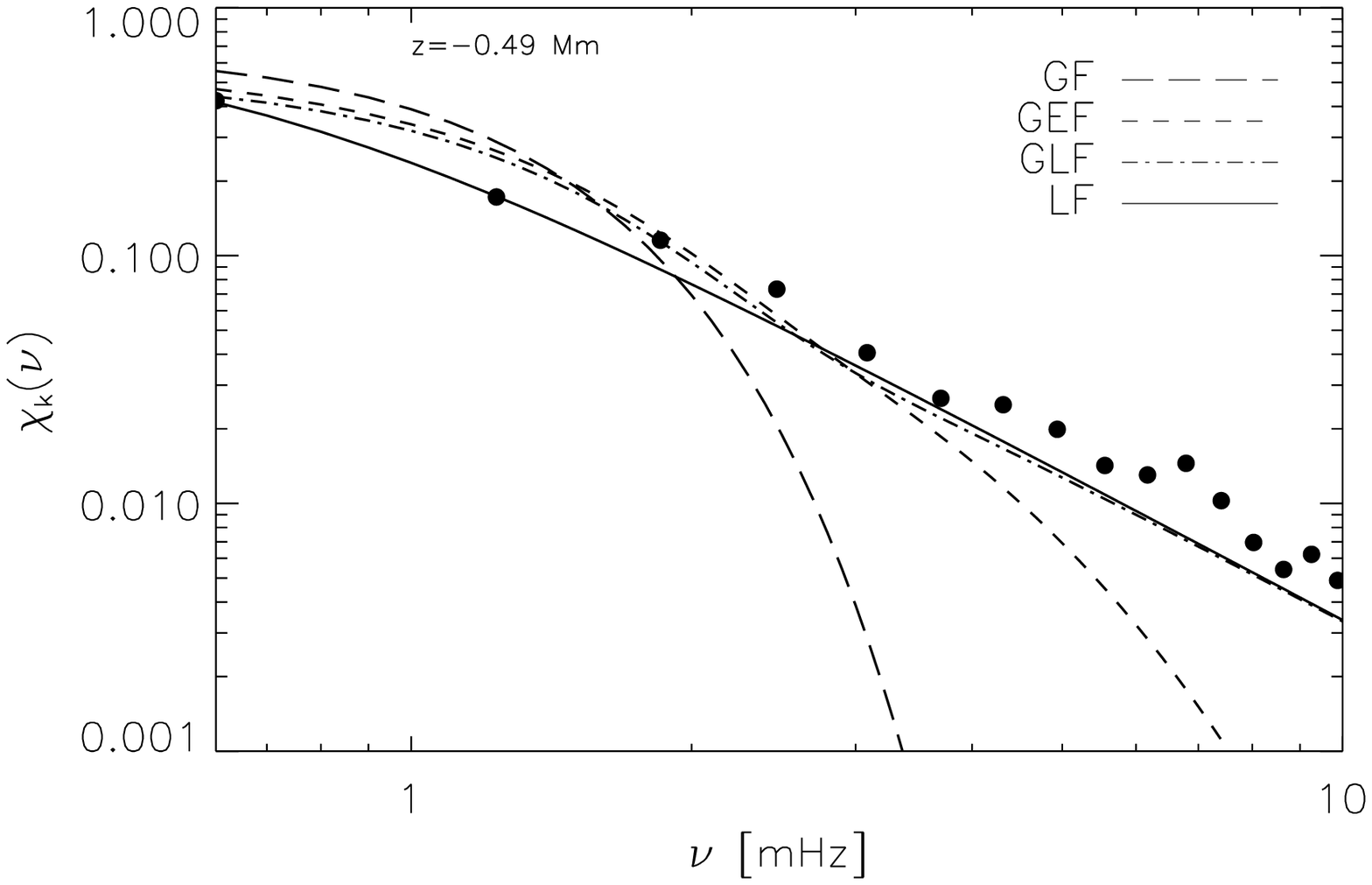}}
       \resizebox{\lenA}{!}{\includegraphics  {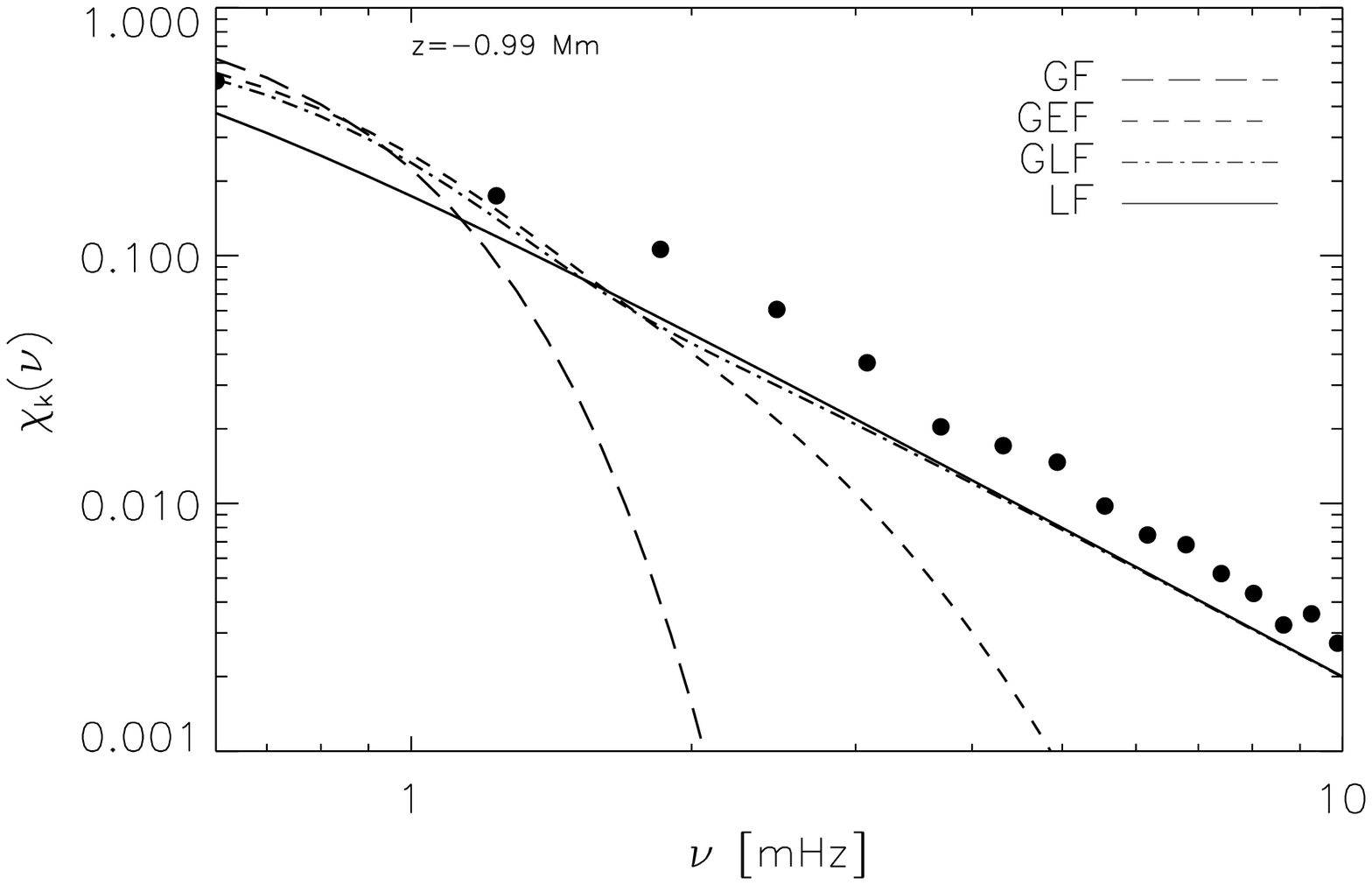}}
        \resizebox{\lenA}{!}{\includegraphics  {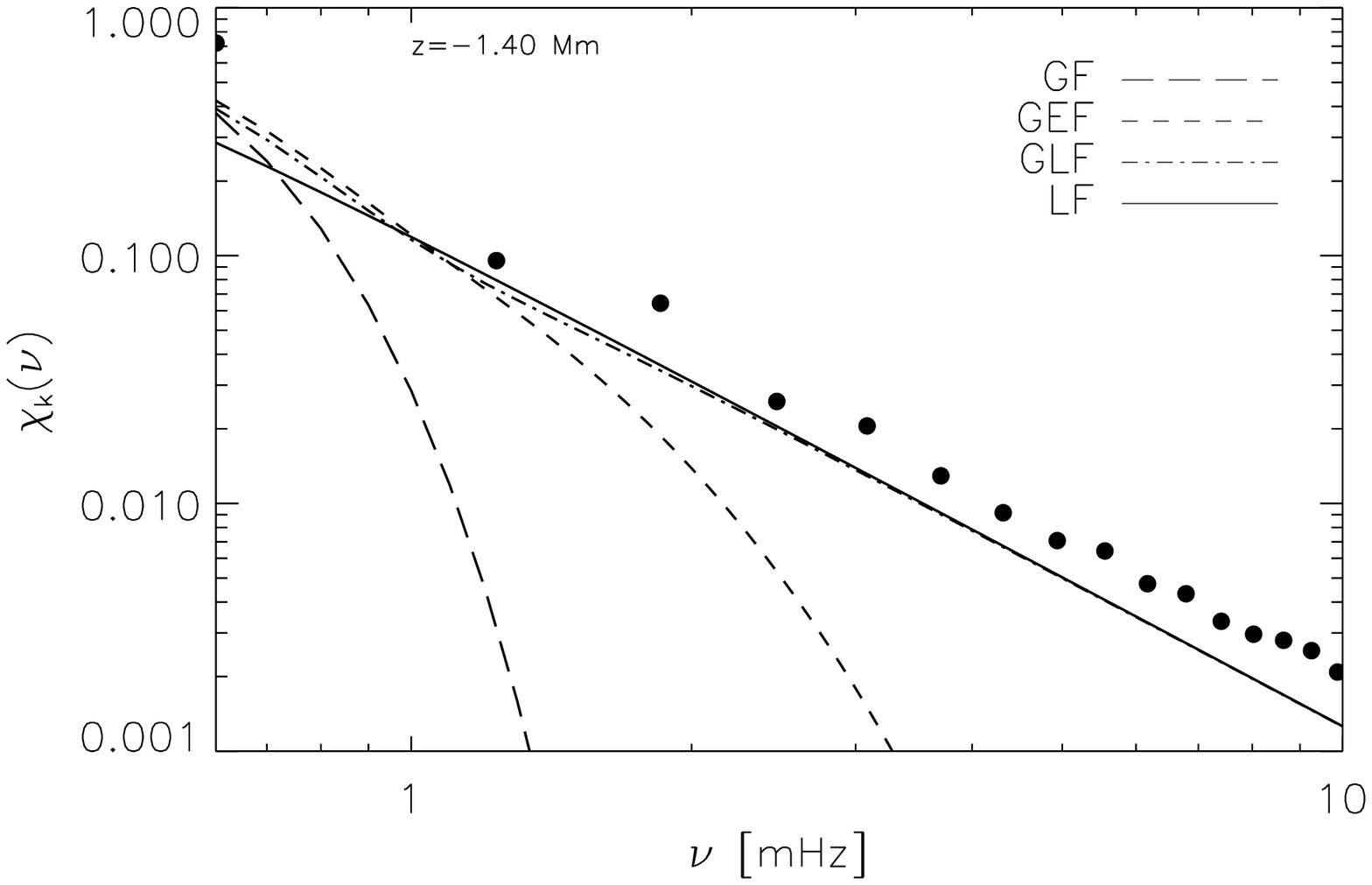}}
\end{center}
\caption{The  filled dots represent 
 $\chi_k(\nu)$ obtained from the simulation for the wavenumber $k_0$  at which $E(k,z)$ peaks.
The solid curves represent the Lorentzian function  (LF, Eq.~\ref{eqn:LF}),
the dots-dashed curves the  Gaussian Lorentzian function (GLF,
Eq.~\ref{eqn:GLF}), the dashed curves the Gaussian Exponential function (GEF, Eq.~\ref{eqn:GEF})
and  the long dashed curves the  Gaussian function (GF, Eq.~\ref{eqn:GF}). 
  In these four  analytical functions,   
$\lambda=1$ is assumed for the calculation of $\omega_k$ (Eqs.~\ref{eqn:tauk}-\ref{eqn:uk2}).
Four different layers are considered: $z=-0.04$~Mm (the top of the
superadiabatic region), 
$z=-0.49$~Mm, $z=-0.99$~Mm and $z=-1.40$~Mm.}
        \label{fig:plot_kwpowerE_lambda1.00}
\end{figure*}

Fig.~\ref{fig:plot_kwpowerE_lambda1.00} presents $\ds \chi_k(\nu)$ 
as it is obtained from the simulation for the wavenumber $k$ at which
$E(k,z)$  peaks ($k=k_0$).
We compare $\ds \chi_k(\nu)$ obtained from the 3D simulation analysis 
with the analytical functions GF, GEF and GLF 
(Eq.~(\ref{eqn:GF}), (\ref{eqn:GEF}), and (\ref{eqn:GLF}) resp.).

At the top of the superadiabatic region (for instance $z=-0.4~$Mm in Fig.1, this is the layer where the excitation is the largest), 
the GF does not correctly model $\chi_k(\nu)$ (see Fig.~\ref{fig:plot_kwpowerE_lambda1.00}). 
However the discrepancies between the GF and the simulation data occur mostly
above  the solar cut-off frequency ($\nu \sim 5.5$~mHz). Discrepancies between 
the GF  and the 3D simulation data have then minor consequences for 
the p~modes excitation in this region.
This is not the case deeper in the simulation where the largest discrepancies  
between the GF and the simulation data occur in the frequency range where the
dominant amount of acoustic energy is injected into the p~modes ($\nu \sim 2 - 4$~mHz).

To reproduce the shape of $\chi_k(\nu)$ obtained with the 3D simulation, one needs 
a function which  at high frequency decreases more slowly than the GF.
For modeling the $\nu$-dependency of $\chi_k(\nu)$, we  thus propose  three analytical functions: 
the Lorentzian function (LF hereafter)
\eqn{
\chi_k (\omega ) =\frac{1} {\pi \omega_k/2} \,\frac{1}{1+ \left( 2 \omega / \omega_k \right )^2} \;  ,
\label{eqn:LF}
}
 the Gaussian  plus an Exponential function  (GEF hereafter) 
\eqn{
\chi_k (\omega ) = {1 \over 2 } \, \left (  \inv  { \omega_k \, \sqrt{\pi}}  e^{-(\omega / \omega_k)^2} 
+ \frac{1}{2 \omega_k}  e ^{-| \omega/\omega_k | } \right ) \;  ,
\label{eqn:GEF}
} and the Gaussian  plus a Lorentzian function (GLF hereafter)
\eqn{
\chi_k (\omega ) = {1 \over 2 } \, \left (  \inv  { \omega_k \, \sqrt{\pi}}  e^{-(\omega / \omega_k)^2} 
+ \frac{1}{\pi \, \omega_k}\, \frac{1}{1+\left (\omega/\omega_k\right)^2} \right ) \; .
\label{eqn:GLF}
}   
All these functions satisfy the condition of normalisation of 
Eq.~(\ref{eqn:chi_omega_norm}).

We first assume a constant $\lambda=1$.
As shown in Fig.~\ref{fig:plot_kwpowerE_lambda1.00}, all these 
non-gaussian functions reproduce the $\nu$-variation of  $\chi_k$
better than that obtained using a GF.

In the middle of the excitation region ($-0.5~{\rm Mm} \lesssim z \lesssim
0.0~{\rm Mm}$)  the overall best agreement is obtained  with the LF. 
Below $ z \sim -0.5~{\rm Mm}$,  the LF does not reproduce $\ds \chi_k$
 well enough but still reproduces its  $\nu$-variation better than the other models. 

\begin{figure*}[ht]
\begin{center}
       \resizebox{\lenA}{!}{\includegraphics  {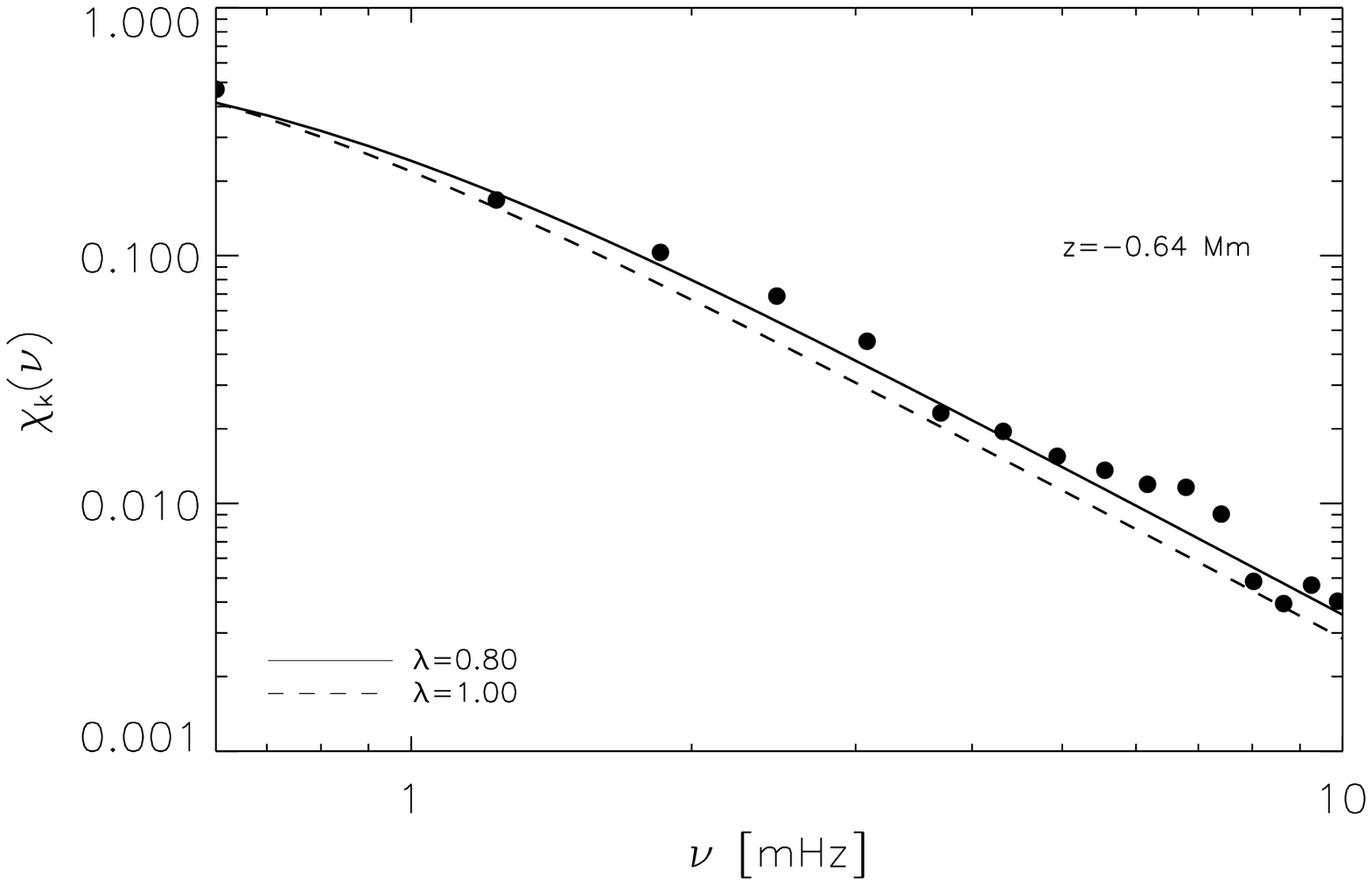}}
        \resizebox{\lenA}{!}{\includegraphics  {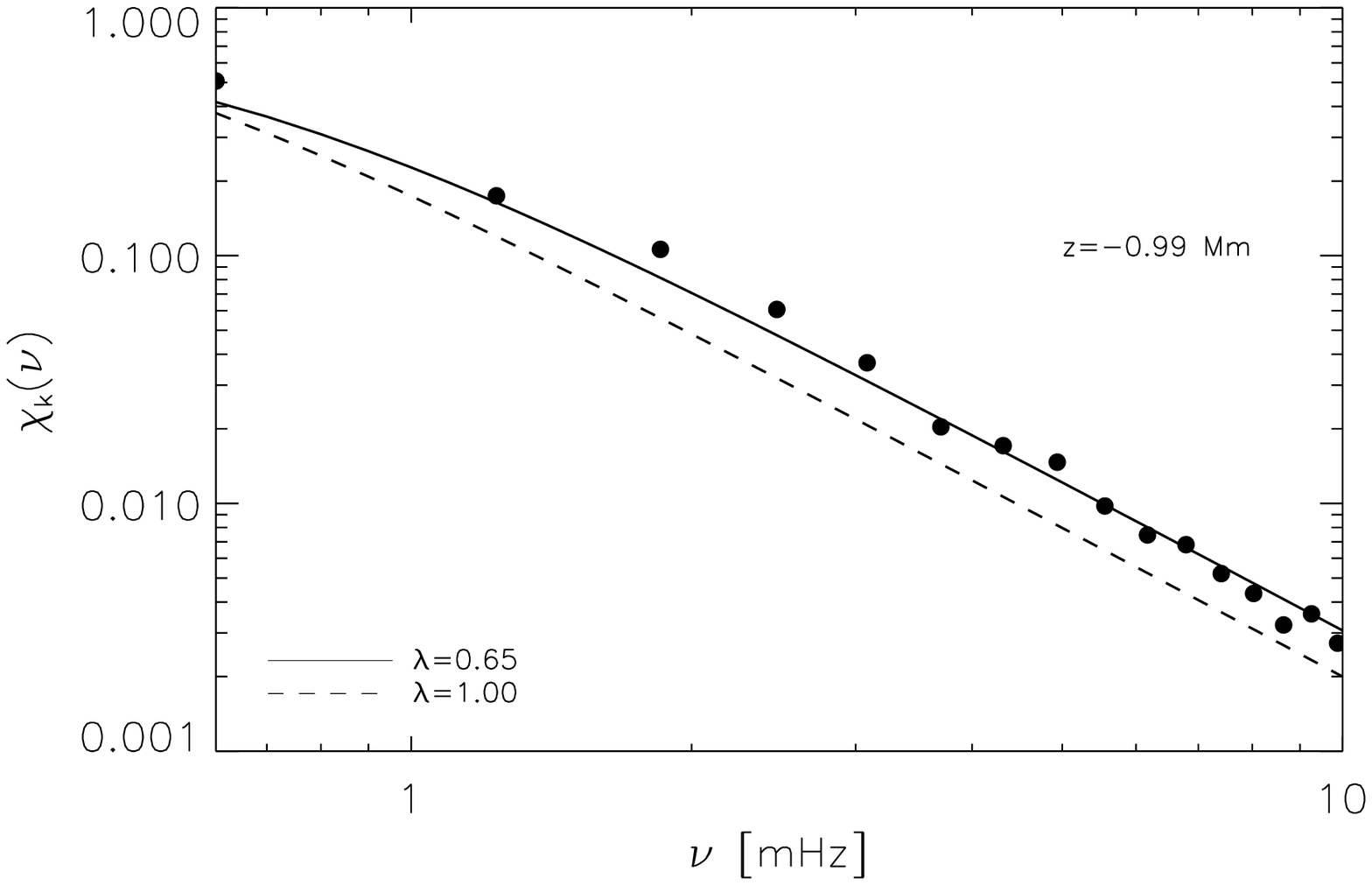}}
\end{center}
\caption{As in Fig.~\ref{fig:plot_kwpowerE_lambda1.00}, the filled dots represent  $ \chi_k(\nu,z)$ 
obtained from the simulation at two different layers : $z=-0.64$~Mm (left panel) and $z=-0.99$~Mm 
(right panel). The other curves represent the LF (Eq.~\ref{eqn:GLF})  with different assumptions for $\lambda$:
The dashed curved correspond to $\lambda=1$ and the solid curves correspond to 
$\lambda=0.80$ at $z=-0.64$~Mm (left panel) and $\lambda=0.65$ at $z=-0.99$~Mm (right panel).}
        \label{fig:plot_kwpowerE_lambda}
\end{figure*}

However   we have so far assumed that $\lambda$ (or equivalently the eddy time
correlation)  is depth independent, which is a
strong assumption. When $\lambda$ is allowed to vary with $z$, 
we find that  decreasing the value of $\lambda$ below $ z \lesssim -0.5~{\rm Mm}$, 
the LF best models $\chi_k$ below $ z \sim -0.5~{\rm Mm}$ 
(e.g. $\lambda=1.6$ at  z=-0.64~Mm and $\lambda \simeq 1.30$ 
at z=-0.99~Mm, see Fig.~\ref{fig:plot_kwpowerE_lambda}).
This shows that
the variation with depth of the characteristic time $\tau_k$ 
(or equivalently the characteristic frequency $\omega_k$) 
is not correctly represented by the relations of
Eqs.~(\ref{eqn:tauk}-\ref{eqn:uk2}) when computed
assuming  a constant $\lambda$   below $ z \sim -0.5~{\rm Mm}$; $\tau_k$
increases faster with depth than expected from the relations 
Eqs.~(\ref{eqn:tauk}-\ref{eqn:uk2}).
It is however found in Sect.~\ref{sec:Consequences in terms of p modes excitation} 
that this feature has negligible effect on $P$.

%(e.g. $\lambda \simeq 0.85$ at z=-0.49~Mm and $\lambda \simeq 0.7$ at z=-0.99~Mm).

\vspace{0.25cm}

The function $\ds \chi_k^z$ also decreases with the frequency more slowly
than the GF (see Fig.~\ref{fig:plot_kwpowerE_lambda1.00}). 
Moreover, decreasing values of $\lambda$  for $z \lesssim -0.5$~Mm provide a
better fit  of  $\ds \chi_k^z$.
But in contrast with  $\chi_k$, 
$\ds \chi_k^z$ is  overall better  modeled with the GEF for $ z \gtrsim -0.5~\textrm{Mm}$
and with the GLF for $z \lesssim   -0.5~\textrm{Mm} $ rather than with the LF (not shown).

\vspace{0.25cm}

We conclude from the frequency analysis  of the 3D simulation 
that the simple gaussian function cannot correctly represent the actual dynamic
properties of  the turbulent medium. 
One may expect that  the GF causes an  underestimation of 
the acoustic energy injected into the solar p~modes.
Instead  the frequency analysis favours a non-gaussian function  
for $\ds \chi_k^z$   and  $\chi_k$ that decreases more slowly 
with $\nu$ than the GF.

%%%%%%%%%%%%%%%%%%%%%%%%%%%%%%%%%%%%%%%%%%
\section{Consequences in terms of p~modes excitation}
\label{sec:Consequences in terms of p modes excitation}

%------------------------------------------------
\subsection{Computations of the excitation rate $P$}

Computation of the excitation rate $P$ is performed as in \citet{Samadi02I}
except that here two analytical functions  other than the GF are assumed for
$\chi_k$, as discussed in
Sect.~\ref{sec:Constraints from the 3D simulation}. 
The computation process is summarised  as follows:
  The eigenfunctions ($\xi_{\rm r}$) and their frequencies ($\nu$) are computed 
with Balmforth's (1992) non-adiabatic code for a solar 1D mixing-length 
model based on Gough's (1977) non-local time-dependent formulation of convection.

The quantities  $\Phi$, $w^2$ and $s^2$ are obtained from the 3D simulation.
 The $k$-dependency of $E(k,z)$ is  the Extended Kolmogorov
 Spectrum  (EKS hereafter)~ defined as ~:
\eqn{
\begin{array}{lccl}
 E(k) \propto (k/k_0)^{+1}   &  \textrm{for} & k_0  >  k >  k_{\rm min}\\
 E(k) \propto  (k/k_0)^{-5/3}  &   \textrm{for} &  k > k_0                \\
\end{array}
\label{eqn:EKS}
}
In Eq.~(\ref{eqn:EKS}), the  wavenumber $k_0$ is the wavenumber at 
which $E(k)$ peaks and $k_{\rm min}$ is the minimal wavenumber 
reached by the 3D simulation  ($k_{\rm min}=1.05~{\rm Mm}^{-1}$). 
The variation with depth of $k_0$ is also given by   the 3D simulation.
 The  $k$-dependency of the EKS reproduces the global features of $E$ arising 
from the 3D simulation. The same model is  considered for  $E_s(k,z)$.
 $E(k,z)$ and $E_s(k,z)$ satisfy the normalisation conditions: \eqn{
\begin{array}{lll}
\vspace{0.2cm}
\ds \int_{k_{\rm min}}^{+\infty}   {\rm d} k  \, E(k,z) & = & {1/2} \, \Phi \, w^2    %& =  & \ds  {1 \over 2} \, < {\bf u}^2- <{\bf u}>^2>(z)
\\ \vspace{0.2cm}
\ds \int_{k_{\rm min}}^{+\infty}   {\rm d} k  \, E_s(k,z)   & =  &  {1/2} \, \tilde s ^2   \; .  %  & =  & \ds {1 \over 2} \, <s2-  <s>^2>(z)
\end{array}
\label{eqn:EEzEs}
}
The total energy contained in  $E(k,z)$ and $E_s(k,z)$, 
and their depth dependencies, are then obtained from the 3D 
simulation according to Eq.(\ref{eqn:EEzEs}). 
These 
theoretical estimates for $P$ are then
compared with the  `observed' $P$ from \citet{Chaplin98}'s seismic 
data, calculated according to the relation :
\eqn{
P (\omega_0) = 2 \eta \, \frac{ I  }{\xi_{\rm r}^2(r_s) } \,  v_s^2 (\omega_0) 
\label{eqn:P_vs2}
}
where    $r_s$ is the radius at which oscillations are measured,   
\eqn{
I \equiv   \int_0^{M} dm \,   \xi_{\rm r}^2 
\label{eqn:I} 
} 
is the  mode inertia and where the mode damping rate ($\eta$) and the mode surface velocity ($v_s$) are obtained from \citet{Chaplin98}.
In Eq.~(\ref{eqn:P_vs2}), the mode mass ${I /  \xi_{\rm r}^2(r_s)}$ is given by  
the GMLT model and we adopt  $r_s = R_\odot + 200$~km  
consistently with \citet{Chaplin98}'s observations.     

%-------------------
\subsection{Comparisons with observations}
\label{sec:Comparisons with observations}

\begin{figure}[ht]
\begin{center}
\resizebox{\hsize}{!}{\includegraphics{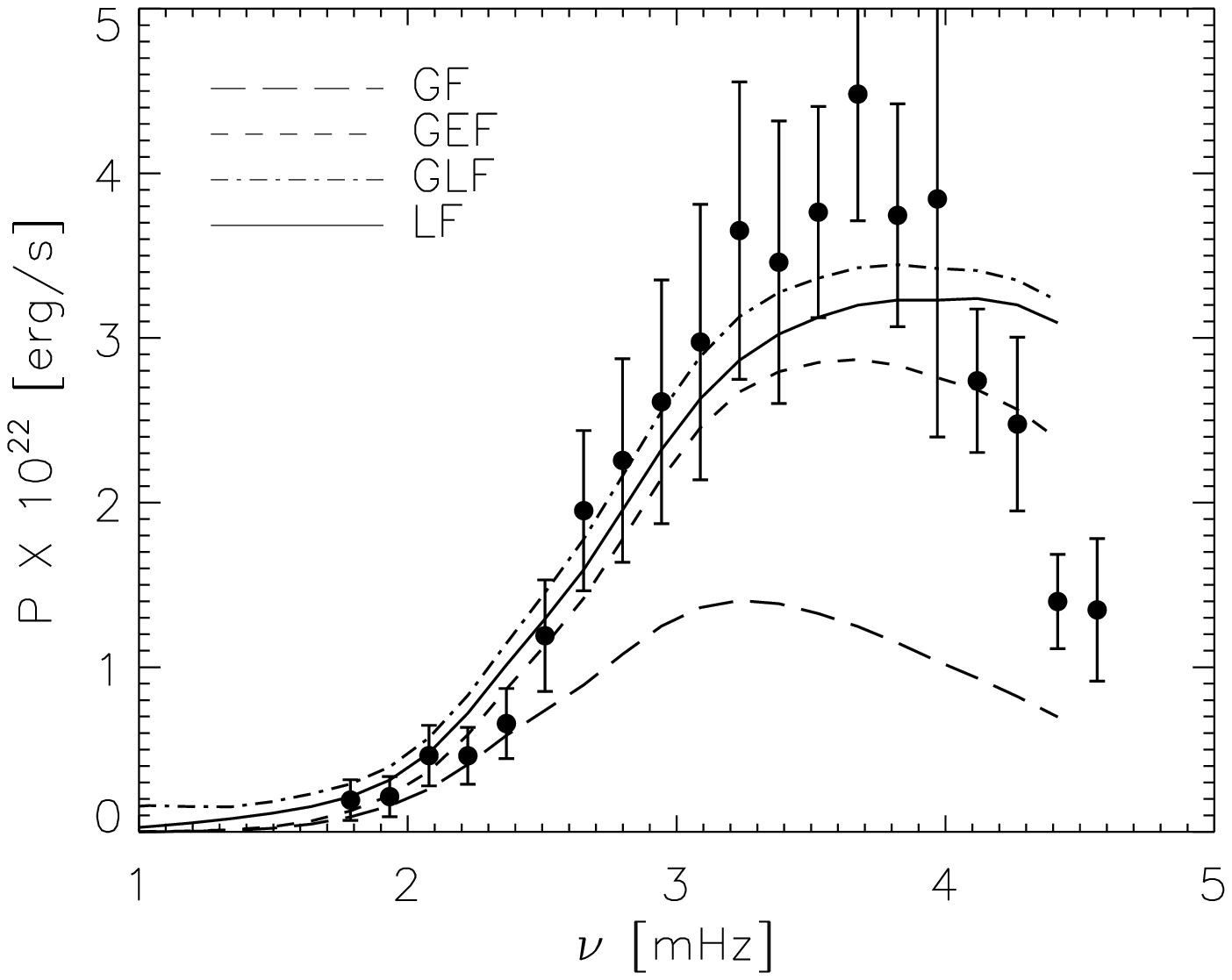}}
\resizebox{\hsize}{!}{\includegraphics{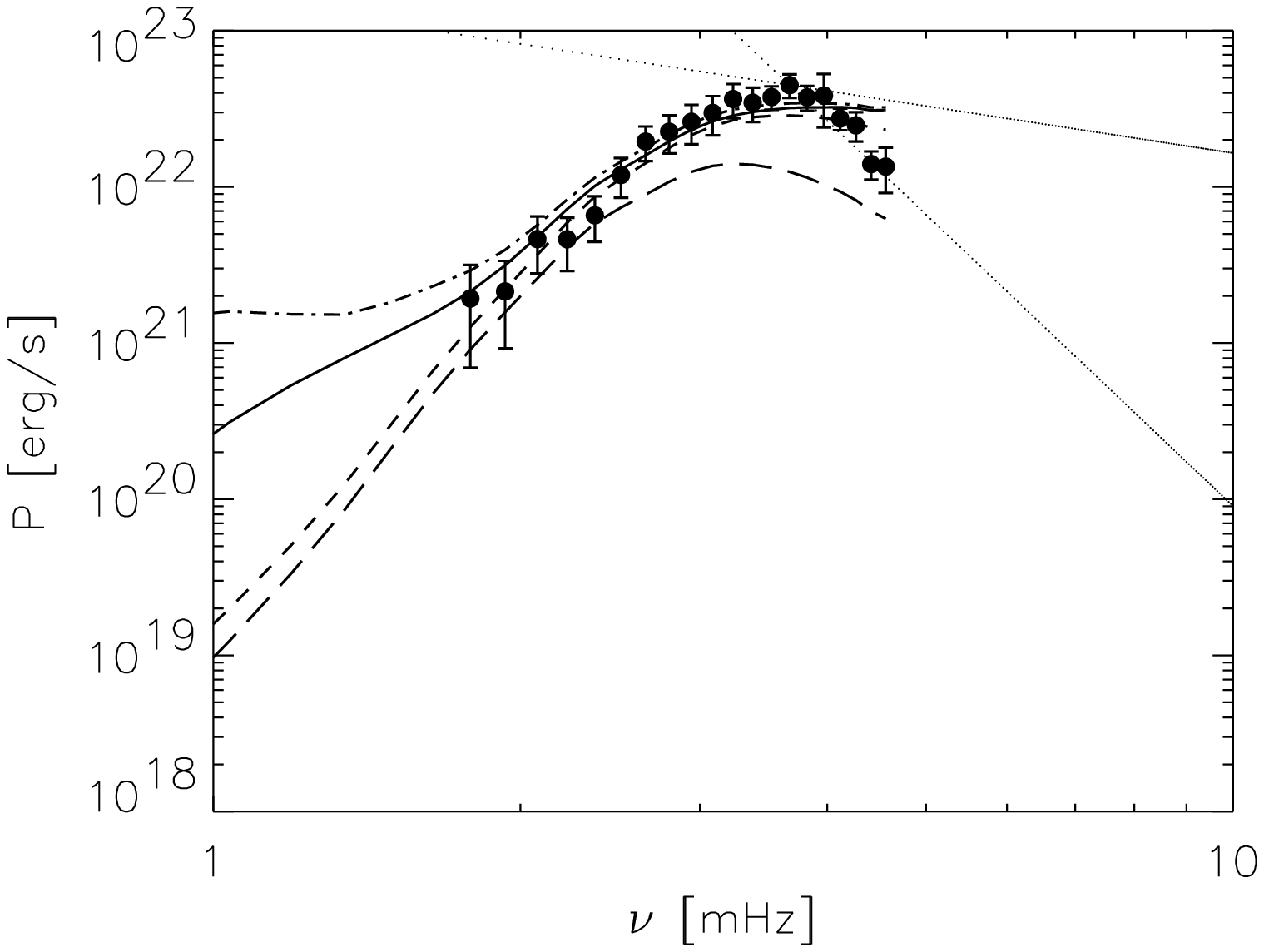}}
\end{center}
\caption{{\bf Top:}  The curves correspond to computed  values of $P(\nu)$ 
obtained with different   analytical functions for $\chi_k(\nu)$: 
the GF (long dashed curve), the GEF (dashed curve),  the GLF (dots-dashed curve) and the LF  (solid curve). 
In all calculations, we assume $\lambda=1$. The dots represent  $P(\nu)$
derived from 
the amplitudes and line widths of the $\ell=0$ p~modes measured by \citet{Chaplin98}.
{\bf Bottom:} same as the top panel but $P$ is plotted in a log-log
representation as it is usually represented in the literature. The vertical
and horizontal scales have been chosen for a easy comparison with equivalent
plots found in \citet{Stein01II}.
The  lines with dots show two different power laws $\nu^{p}$~: 
one with $p=-6.2$ and the other with $p=-1$.
}
\label{fig:Poscgh_chiw}
\end{figure}

We  investigate the effect of using  different analytical functions for $\chi_k$ 
(Sect.~\ref{sec:Constraints from the 3D simulation}) in the computation of $P$ .
%However as in Paper~I and for the sake of simplicity we assume that the frequency component of $E_s(k,\omega)$ is the same than this of the turbulent velocity field $E(k,\omega)$.
We first assume  a constant value  $\lambda=1$. 
Results  are shown in Fig.~\ref{fig:Poscgh_chiw}.
Computations performed with the GF underestimate the
observed $P$ values  by a factor $\sim 2.7$.
On the other hand,  the LF, GEF and GLF choices result in larger values for the computed $P$  
than the GF one  ($\sim 2$ times larger). This brings
them closer to the observations, compared with the GF choice for $\chi_k$.
The reason is that all the non-gaussian functions 
(the LF , the GEF and the GLF)
 -~ which indeed better model  $\chi_k^z$ and $\chi_k$  from the 3D 
simulation than does the GF~-  
 decrease more slowly with $\nu$ than the GF in the  frequency
range where the mode amplitudes are large ($\nu \simeq 2 - 4$~mHz). 
Consequently a larger amount of acoustic energy is injected into the modes with the non-gaussian functions than with the GF.

\vspace{0.2cm}

In Sect.~\ref{sec:Inferred properties of chikz}, 
the overall best models for $\chi_k$ %and for $\ds \chi_k^z$ 
  were obtained with the LF and with decreasing values of $\lambda $  below $z \sim -0.5$~Mm.
We use a simple model for the depth variation of $\lambda$:
\eqn{
\begin{array}{llll}
\lambda & = 1                         & \textrm{for} & z > -0.5~{\rm Mm}
\\\lambda & = 0.9 + 0.71  (0.64 \textrm{Mm} +z )  & \textrm{for} &  -0.5 \ge z \ge -1~{\rm Mm}
\\\lambda & = 0.35                      & \textrm{for} & z < -1~{\rm Mm}  
\end{array}     
\label{eqn:lambda_vs_z}
}
%we assume  $\lambda=0.9$ for  $z <-0.5$~Mm and $\lambda=1$ for $z \geq -0.5$~Mm.
We  have computed $P$ according to the simple model of Eq.~(\ref{eqn:lambda_vs_z}) and assuming 
the  LF. We find no significant changes for $P$   compared to the calculations in which a constant 
value  $\lambda=1$ is assumed  (not shown).

\vspace{0.2cm}

In Sect.~\ref{sec:Constraints from the 3D simulation} we found that $\ds \chi_k^z$ is   better  
modelled with   the GEF for $ z \gtrsim -0.5~\textrm{Mm}$  and with the GLF for $z \lesssim  
 -0.5~\textrm{Mm} $ rather than with the LF. However, as the stochastic excitation is the largest 
 in the range $-0.5~\textrm{Mm}  \lesssim z \lesssim 0$~Mm,  we can assume the GEF in all the domain.
The LF results  in  a value for $P_{\rm max}$   slightly larger  than  the one
 resulting from  the  GEF (only $\sim 1.2$ larger).  
A better agreement is then  obtained with the analytical functions 
which fits best $\chi_k$ (i.e. the LF)  than the one which fits best  $\ds
\chi_k^z$ (i.e. the GEF)
in contrast with the intuitive idea mentioned in Sect.~\ref{sec:The dynamic
model of turbulence}
that the excitation of  radial p~modes  depends rather on the properties of $\ds
\chi_k^z$   than  on those of  $\chi_k$.  

In the frequency range where observational constraints  are available, 
 differences between results 
obtained with the different adopted non-gaussian functions are  
of the same order as the actual error 
bars associated with the observed $P$ values.

Below the frequency range of the observations  -~ i.e. below $\nu \lesssim 1.8~$mHz ~ - the differences 
between the different non-gaussian functions are very large compared to the error bars (see bottom panel of Fig.~\ref{fig:Poscgh_chiw}).
Those differences are  related  to differences in the  $\nu$-variation of the non-gaussian models.
Observational constraints at low frequency could therefore confirm 
that the LF is indeed 
the best representation for  $\chi_k(\nu)$.

\vspace{0.2cm}

Important discrepancies still remain at high frequency ($\nu \gtrsim 3.5$~mHz).
The excitation rate derived from the observations decreases 
as $\sim \nu^{-6.2}$ above $\nu \simeq 3.5$~mHz) 
whereas the computed $P$ decreases as $\sim \nu^{-1}$ 
(see bottom panel of Fig.~\ref{fig:Poscgh_chiw}). 
Possible origins of this discrepancy are discussed 
in Sect.~\ref{sec:Possible origin of the remaining discrepancy}.

\begin{figure}[ht]
\begin{center}
\resizebox{\hsize}{!}{\includegraphics{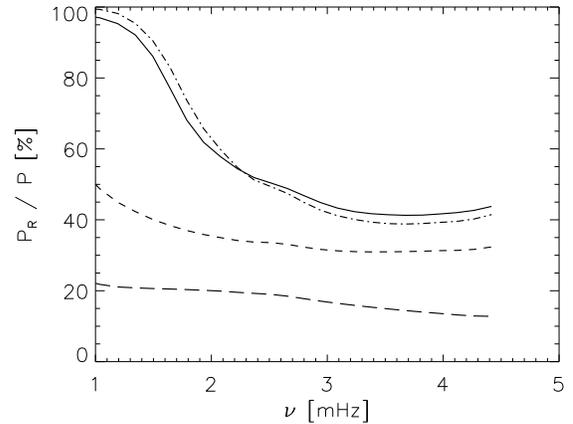}}
\end{center}
\caption{ {\bf Top:} 
Same as Fig.~\ref{fig:Poscgh_chiw} for the relative contribution of 
the Reynolds stress, $P_{R}$ to the total acoustic energy $P$.
 }
\label{fig:Poscgh_chiw_PrsP}
\end{figure}

\vspace{0.5cm}

Another consequence of a non-gaussian  dependence of $ \chi_k$ (or
$\ds \chi_k^z$) with the frequency is a larger  relative  contribution of the Reynolds stress $P_{R}$ 
to the mode excitation rate $P$. This is shown in  Fig.~\ref{fig:Poscgh_chiw_PrsP}.
The GF generates a relative contribution of the  Reynolds stress to the
excitation ($P_R/P$) which is smaller than that obtained assuming a non-gaussian function  
(e.g. for the LF, the relative  contribution of $P_R$ to the excitation is at least 
$\sim 2$ times larger than with the GF).
Excitation by the entropy fluctuations takes place predominantly at the top of
the excitation  region over a thin layer ($\lesssim 0.2$~Mm) while that due
to the Reynolds stress  extends  deeper below ($\sim 0.5 - 2$~Mm).
At the top of the excitation region, the discrepancy between the GF and $\chi_k$
inferred from the 3D  simulation mainly occurs above $\nu \simeq 5$~mHz and
thus has a  small impact on mode excitation.
This is not the case deeper in the excitation region where the GF
under-estimates $\chi_k$ in a  frequency range increasing inward.

%%%%%%%%%%%%%%%%%%%%%%%%%%%%%%%%%%%%%%%%%%
%\newpage
\section{Conclusion and discussion}
\label{sec:Conclusion}

\subsection{A Non-Gaussian eddy time correlation}

In the present work we characterize \emph{empirically}  $\chi_k$ and
of  $\chi_k^z$,  the frequency components of the  correlation product  of the turbulent velocity field
and of its  vertical component respectively. 
%the dynamic models of turbulence respectively 
% associated with $E(k,\omega,z)$,  the turbulent  kinetic energy spectrum and
% with $E_z(k,\omega,z)$, the vertical component of the  turbulent kinetic energy spectrum.
A frequency analysis of a solar 3D simulation shows that at large scales ($k \sim k_0$)
the gaussian function significantly underestimates   $\chi_k$ and   $\chi_k^z$
in the  frequency range  ($\nu \simeq 2 - 4$~mHz) 
where acoustic energy injected into the solar p~modes is the largest.

% Assuming the GF,

 As a result, the maximum value of $P$ is  found $\sim 2.7$ smaller than the 
solar seismic constraints.
%  This is because the GF overestimates $\chi_k$ and $\chi_k^z$  in most part of
%  the frequency 
%range where p~modes have large amplitudes.

This partly explains the \emph{underestimate} of the values of 
solar p~mode excitation rates
obtained 
by \citet{Houdek99} whose computations are based on the theoretical expression by \citet{Balmforth92c}
and  by \citet{Samadi02I}.

In order to reproduce   the main properties of $\chi_k$ (or  $\chi_k^z$)~,  
one has to consider a model  which must decrease more slowly with $\nu$  than 
the GF.
We then assume  for $\chi_k$ and $\chi_k^z$   three different   simple analytical forms  :
the Lorentzian Function (LF), the so-called 'Gaussian Exponential Function' 
(GEF, which is composed by the GF plus an exponential function) and 
the so-called 'Gaussian Lorentzian Function' (GLF, which is composed by the GF plus a Lorentzian  function).

From the top of the excitation region (which corresponds to the top of the
superadiabatic region)  down to the middle of the excitation region ($z \sim -0.5$~Mm where $z$ is the distance to the radius at the photosphere), 
the best agreement between  $\chi_k$ and the  analytical approximations    is
obtained with  the LF and with $\lambda=1$. 
Deeper  within the excitation region ($z \lesssim -0.5~$Mm), 
the  agreement is  better for  $\lambda <1$ and $\lambda$  decreasing with depth.

The frequency dependencies of $\chi_k^z$ and $\chi_k$ are found to be very similar. 
However $\ds \chi_k^z$ is best modeled by the GEF.
As for  $\chi_k$, the  agreement is  better below $z \sim -0.5$~Mm  
with decreasing values of the parameter $\lambda$ than with  $\lambda=1$.  

Assuming a  non-gaussian function  -~either the LF, the GEF or the GLF ~-
results in  values for $P_{\rm max}$, the  maximum of  excitation power,
which are $\sim 2$ times larger than when assuming the GF 
and brings $P_{\rm max}$ much closer to the  maximum of  $P$ derived from the 
solar seismic data of \citet{Chaplin98}. 

We also find that taking into account the variation of  $\lambda$ with depth 
for $z$ below $  -0.5$~Mm does not significantly  change the values of $P$. 
A constant value can then be assumed in the calculation of 
the solar p~mode excitation rates. The constant value of $\lambda$ on the
other hand plays an important role and we find $\lambda=1$.  

We have investigated the sensitivity to the adopted representation for $\chi_k$ : 
Although the LF fits best the  $\nu$-variation of $\chi_k$ inferred from the 3D simulation, 
the GLF results in value for $P_{\rm max}$ closer to the seismic constraints.
However, the differences obtained with the different non-gaussian approximations 
for $\chi_k$  
are globally smaller than the actual error bars associated with the observations 
of \citet{Chaplin98}.  
On the other hand, below the frequency range where observational constraints on $P$ are available 
(i.e. below $\nu \lesssim 1.8~$mHz),  the differences  between $P$ obtained with different 
non-gaussian functions are very large compared to the current  error bars. Those  differences 
are directly related to the diffences in the $\nu$-variation of the  non-gaussian forms investigated in this work.
This suggests that accurate enough data below this frequency range, could provide   
confirmation that the LF is indeed the best model for  $\chi_k$.

                        %-----%
\subsection{Relative contribution of the entropy fluctuations to the excitation}

%The nature and the general properties of the source term of excitation
%due to the Reynolds tensor are well established \citep{GK77}.
%However during last decade the source of excitation that originates in the fluctuations of the entropy  gave rise to major controversies \citep[e.g.][]{Balmforth92c,GMK94}.

The non-gaussian character  of $\chi_k$ causes the excitation
region to extend  deeper ($\sim 500$~km for modes of order $n=20$)  
than with the GF ($\sim 200$~km resp.).
The largest  entropy fluctuations  mainly occur  at the outermost part of the 
convective zone (CZ) over a very thin region ($\sim 100$~km) while excitation
by the  Reynolds stress contribution occurs on a more extended region.
Consequently the non-gaussian   property  of $\chi_k$ leads to a relatively larger
contribution of the  Reynolds stress to the excitation than in the case
of a GF.
As a result, the  Reynolds stress  contribution is of the same order as the contribution 
arising from  the advection of the turbulent fluctuations of entropy by the 
turbulent movements (the so-called entropy source term).
This is in contrast with  previous results \citep{Samadi00II} based on the GF
which concluded that the entropy source term dominates the Reynolds stress by about $\sim 20$.
It also differs  with results by \citet{GMK94} who found that the excitation 
arising directly from the entropy fluctuations dominates by about $\sim 10$.

On the other hand, in \citet{Stein01II}, the excitation by turbulent pressure
 (Reynolds stress)  is found dominant ($\sim 4$ times larger) whereas here 
we find  that the  contribution of the entropy source term cannot be neglected.
Whether this is the signature of some deficiency in the present excitation model is an open
 question. 

%------------------------------------------------------
\subsection{Summary}

We show that the usually  adopted  \emph{gaussian function}  for  $\chi_k$ is 
neither consistent with the  properties of  $\chi_k$ inferred from the 3D 
simulation nor does it reproduce the observed  maximum of the solar p-modes 
excitation rates.

Following an empirical  approach we improve  the model  of the convective eddy 
time-correlation  $\chi_k$ which enters the current model of stochastic excitation. 
We then show that to reproduce \emph{both}  the $\nu$-variation  of  $\chi_k$ as 
inferred from the 3D simulation  and the observed  maximum of the solar p-modes 
excitation rates one has to consider  a non-gaussian form which decreases at high 
frequency slower than the GF, as do the different non-gaussian functions investigated 
here.

The use of  non-gaussian functions, for instance the LF, reproduces 
reasonably well the maximum value of the rate at which solar p-modes are excited  
\emph{without any ajdustements of free parameters}  or \emph{without introducing a 
scaling factor}, in contrast with previous approaches 
\citep[e.g.][]{Balmforth92c,GMK94,Samadi00II}. We then solve the problem of the  
underestimation by the previous theoretical approaches. Furthermore the use of 
such a non-gaussian form for $\chi_k$  makes the  contribution of the turbulent 
pressure to the excitation much larger than in previous works making our results 
more consistent with that by \citet{Stein01II}.

Our investigation  clearly   emphasizes  the \emph{non-gaussian character of the 
solar p~modes excitation}  as a  result  of the \emph{non-gaussian property of the 
convective eddies time-correlations}. It also  shows that  the dynamic properties 
of the solar turbulent convection inferred from the 3D simulation are consistent 
with the helioseismic data.

\vspace{0.2cm}

We stress that only simple non-gaussian forms for $\chi_k$ have been 
investigated here. More sophisticated forms are likely to
improve the agreement with the  $\nu$-dependency of  $\chi_k$ (or $\chi_k^z$).  
This would not affect the main conclusions presented in the present paper.

%------------------------------------------------------
\subsection{Possible origin of the non-gaussian property of $\chi_k$}

We recall that $\chi_k$  measures the temporal evolution of the correlation between two 
points of the turbulent medium separated by a distance of $\sim 2\pi/k$.
A gaussian  time-correlation means that the fluid motions in the medium are random in time.
Departure from a gaussian time-correlation  at large scales ($k \sim k_0$)
suggests that a strong correlation exists at that  scale.

 Downward plumes are likely to be responsible for the non-gaussian 
 behaviour  of $\chi_k$. Downward and upward
 convective motions are indeed  highly  asymmetric \citep{Stein98}: downward flows are
 associated  with  patterns (plumes) which are more coherent  than
 the upward moving structures \citep{Rieutord95}. The upward flows are associated with 
 less coherent and more random structures (granules)  characterised by 
a broad variety of sizes and  lifetimes \citep{Rieutord95}. 
The non-gaussian behavior of $\chi_k$  can most probably  be attributed to plumes. 
This however remains to be checked (work in progress).

%------------------------------------------------------
\subsection{Possible origin of the remaining discrepancy}
\label{sec:Possible origin of the remaining discrepancy}

Despite a clear improvement in the agreement between observed and theoretical
excitation rates,  important discrepancies between the computed $P$ and  the solar
measurements still remain
 at high frequency $\nu \gtrsim 3.5$~mHz 
(see Sect.~\ref{sec:Comparisons with observations} and Fig.~\ref{fig:Poscgh_chiw}).

On the `observational side',  at high frequency,
larger uncertainties for  the damping rates $\eta$ 
induce larger uncertainties on the derived supply energy rates.

On the theoretical side, part of the discrepancy might well be attributed  to 
a poor description of the eigenfunctions at high frequency.
Indeed, the discrepancies between the calculated  eigenfrequencies and the
observed ones are largest at high frequency ($\nu \gtrsim 3~$mHz).
This indicates that the description of the eigenfunctions are less 
accurate at high frequency. 
As the expression for calculating $P$ involves the first and second derivatives 
of the mode eigenfunction, the lack of accuracy in the calculation of the
eigenfunctions  has a larger impact  on $P$ at high frequencies  
than at small frequencies.

Other possible causes can  perhaps be  related to our simplified excitation 
model which assumes isotropic turbulence.
Indeed the current theory assumes that the stochastic excitation is the same 
in all three directions, particularly  between the ascending and descending flows.
However the kinetic energy and entropy fluctuations are larger in the  downward
flows than in the upward flows \citep{Stein98}.
Therefore the driving  arising from  the advection of the turbulent
fluctuations of entropy by the turbulent movements  differs significantly between
the elements moving downwards and those moving upwards.
As the entropy fluctuations are largest in the outermost part of the
convective zone, the above mentioned asymmetry will predominantly affect  the high
frequency modes.

Moreover, it is also assumed that  the total kinetic energy, $E$, is
isotropically  injected in all 3 directions. Excitation of the radial p~modes
results  from the vertical component of the velocity.
However at the top of the convective zone, the distribution of kinetic energy
in  $E(k,z)$  and  in  $E_z(k,z)$ are very different  from each other.
These differences may affect more strongly  the high frequency modes.
Consequences of these departures  from the isotropic assumption need to be
further investigated.

\subsection{Perspectives}

The non-gaussian property of $\chi_k$  and its consequences for 
the stochastic excitation has been investigated 
 so far only for  the \emph{Sun}. However such a  non-gaussian feature 
of the turbulence will most likely also be of importance for 
solar-like oscillating stars more
massive than the Sun, provided our analysis is also valid for these stars. 
This can 
 substantially  change  the  excitation spectrum $P$ 
 for such stars compared to that which is currently predicted.

Therefore  investigations of p~mode excitation in hotter and 
more massive stars must be undertaken, which  should  proceed in two steps :
first, the validity  of the present results obtained in the solar case 
must be investigated for other stars  with, for instance,  the help of dedicated 
3D  simulations.
The  conclusions which will drawn from this first step 
must be used  in a second step to study the 
frequency dependence and the magnitude of $P$  
for different  solar-like oscillating stars 
\citep[see preliminary results in ][]{Samadi02c}.

Future space missions such as COROT 
\citep{Baglin98}, MOST \citep{Matthews98} and 
Eddington \citep{Favata00} will provide high-quality data on seismic 
observations. COROT will be the first mission that will provide 
high precision mode amplitudes and linewidths in other stars.
This high-quality data will allow us to derive the excitation rate $P$.
and will provide improved observational constraints on the theory 
of stochastic excitation which is, at present, poorly constrained
by observation.

%%%%%%%%%%%%%%%%%%%%%%%%%%%%%%%%%%%%%%%%%%
\begin{acknowledgements}
We thank H.-G. Ludwig for useful discussions and valuable help in analysing the 
simulated data and we are indebted to G. Houdek for providing us the 1D solar model.
We gratefully thank our referee (T. Appourchaux) for pertinent remarks 
which led to an improvement of the paper.
RS's work has been supported in part by the Particle Physics and 
Astronomy Research Council of the UK under grant PPA/G/O/1998/00576.
\end{acknowledgements}

%--------------------------------------------------------------------
% BIBLIOGRAPHY
%--------------------------------------------------------------------

%\bibliography{../../biblio}

\begin{thebibliography}{22}
\expandafter\ifx\csname natexlab\endcsname\relax\def\natexlab#1{#1}\fi

\bibitem[{{Baglin} \& {The Corot Team}(1998)}]{Baglin98}
{Baglin}, A. \& {The Corot Team}. 1998, in IAU Symp. 185: New Eyes to See
  Inside the Sun and Stars, Vol. 185, 301

\bibitem[{{Balmforth}(1992)}]{Balmforth92c}
{Balmforth}, N.~J. 1992, \mnras, 255, 639

\bibitem[{Batchelor(1970)}]{Batchelor70}
Batchelor, G.~K. 1970, The theory of homogeneous turbulence (University Press)

\bibitem[{{Chaplin} {et~al.}(1998){Chaplin}, {Elsworth}, {Isaak}, {Lines},
  {McLeod}, {Miller}, \& {New}}]{Chaplin98}
{Chaplin}, W.~J., {Elsworth}, Y., {Isaak}, G.~R., {et~al.} 1998, \mnras, 298,
  L7

\bibitem[{{Favata} {et~al.}(2000){Favata}, {Roxburgh}, \&
  {Christensen-Dalsgaard}}]{Favata00}
{Favata}, F., {Roxburgh}, I., \& {Christensen-Dalsgaard}, J. 2000, in The Third
  MONS Workshop : Science Preparation and Target Selection, 49--54

\bibitem[{{Goldreich} \& {Keeley}(1977)}]{GK77}
{Goldreich}, P. \& {Keeley}, D.~A. 1977, \apj, 212, 243

\bibitem[{{Goldreich} {et~al.}(1994){Goldreich}, {Murray}, \& {Kumar}}]{GMK94}
{Goldreich}, P., {Murray}, N., \& {Kumar}, P. 1994, \apj, 424, 466

\bibitem[{{Gough}(1977)}]{Gough77}
{Gough}, D.~O. 1977, \apj, 214, 196

\bibitem[{{Houdek} {et~al.}(1999){Houdek}, {Balmforth},
  {Christensen-Dalsgaard}, \& {Gough}}]{Houdek99}
{Houdek}, G., {Balmforth}, N.~J., {Christensen-Dalsgaard}, J., \& {Gough},
  D.~O. 1999, \aap, 351, 582

\bibitem[{{Matthews}(1998)}]{Matthews98}
{Matthews}, J.~M. 1998, in Structure and Dynamics of the Interior of the Sun
  and Sun-like Stars, 395

\bibitem[{{Musielak} {et~al.}(1994){Musielak}, {Rosner}, {Stein}, \&
  {Ulmschneider}}]{Musielak94}
{Musielak}, Z.~E., {Rosner}, R., {Stein}, R.~F., \& {Ulmschneider}, P. 1994,
  \apj, 423, 474

\bibitem[{{Osaki}(1990)}]{Osaki90}
{Osaki}, Y. 1990, in Lecture Notes in Physics : Progress of Seismology of the
  Sun and Stars, ed. Y.~{Osaki} \& H.~{Shibahashi} (Springer-Verlag), 75

\bibitem[{{Rieutord} \& {Zahn}(1995)}]{Rieutord95}
{Rieutord}, M. \& {Zahn}, J.-P. 1995, \aap, 296, 127

\bibitem[{{Samadi}(2001)}]{Samadi01}
{Samadi}, R. 2001, in SF2A-2001: Semaine de l'Astrophysique Francaise, E148
  (astro--ph/0108363)

\bibitem[{{Samadi} {et~al.}(2003){Samadi}, { Nordlund}, {Stein}, {Goupil}, \&
  {Roxburgh}}]{Samadi02I}
{Samadi}, R., { Nordlund}, {\AA}., {Stein}, R., {Goupil}, M.-J., \& {Roxburgh},
  I. 2003, \aap (in press)

\bibitem[{{Samadi} \& {Goupil}(2001)}]{Samadi00I}
{Samadi}, R. \& {Goupil}, M.~. 2001, \aap, 370, 136 (Paper~I)

\bibitem[{{Samadi} {et~al.}(2001){Samadi}, {Goupil}, \&
  {Lebreton}}]{Samadi00II}
{Samadi}, R., {Goupil}, M.~., \& {Lebreton}, Y. 2001, \aap, 370, 147

\bibitem[{{Samadi} {et~al.}(2002){Samadi}, {Nordlund}, {Stein}, {Goupil}, \&
  {Roxburgh}}]{Samadi02c}
{Samadi}, R., {Nordlund}, A., {Stein}, R.~F., {Goupil}, M.-J., \& {Roxburgh},
  I. 2002, in SF2A-2002: Semaine de l'Astrophysique Francaise, meeting held in
  Paris, France, June 24-29, 2002, Eds.: F. Combes and D. Barret, EdP-Sciences
  (Editions de Physique), Conference Series

\bibitem[{{Stein}(1967)}]{Stein67}
{Stein}, R.~F. 1967, Solar Physics, 2, 385

\bibitem[{{Stein} \& {Nordlund}(1998)}]{Stein98}
{Stein}, R.~F. \& {Nordlund}, A. 1998, \apj, 499, 914

\bibitem[{{Stein} \& {Nordlund}(2001)}]{Stein01II}
{Stein}, R.~F. \& {Nordlund}, {\AA}. 2001, \apj, 546, 585

\bibitem[{{Tennekes} \& {Lumley}(1982)}]{Tennekes82}
{Tennekes}, H. \& {Lumley}, J. 1982, A first course in turbulence, $8^{th}$
  edn. (MIT Press)

\end{thebibliography}
%\input samadi3D2.bbl
\bibliographystyle{aa}

\end{document}